\documentstyle[preprint,prl,aps]{revtex}
\begin{document}
\def\sn2{$\sin^22\theta$}
\def\dm2{$\Delta m^2$}
\def\ch2{$\chi^2$}
\def\ltap{\ \raisebox{-.4ex}{\rlap{$\sim$}} \raisebox{.4ex}{$<$}\ }
\def\gtap{\ \raisebox{-.4ex}{\rlap{$\sim$}} \raisebox{.4ex}{$>$}\ }
\newcommand{\EQ}{\begin{equation}}
\newcommand{\EN}{\end{equation}}
\tightenlines
\draft
\begin{titlepage}
\preprint{\vbox{\baselineskip 10pt{
\hbox{Ref. SISSA 45/97/EP}
\hbox{April 1997}}}}
\vskip -0.4cm
\title{ \bf Describing Analytically the Matter-Enhanced
Two-Neutrino Transitions in a Medium }
\author{S.T. Petcov $^{a,b)}$\footnote{Also at:
Institute of Nuclear Research and Nuclear Energy, Bulgarian Academy
of Sciences, BG--1784 Sofia, Bulgaria.}}
\address{a) Scuola Internazionale Superiore di Studi Avanzati, I-34013
Trieste, Italy}
\address{b) Istituto Nazionale di Fizica Nucleare, Sezione di Trieste, I-34013 
Trieste, Italy}  
\maketitle
\begin{abstract}
\begin{minipage}{5in}
\baselineskip 16pt
A general exact analytic expression for 
the probability of matter-enhanced
two-neutrino transitions in a medium 
(MSW, RSFP, generated by neutrino FCNC 
interactions, etc.) is derived. The probability is expressed 
in terms of three real functions of the parameters 
of the transitions: the ``jump''
probability and two phases (angles). The results obtained 
can be utilized, in particular, in the studies 
of the matter-enhanced transitions/conversions 
of solar and supernova neutrinos. 
An interesting similarity between the Schroedinger 
equation for the radial
part of the non-relativistic wave function of the hydrogen atom
and the equation governing the MSW transitions of solar 
neutrinos in the exponentially varying matter density in the 
Sun is also briefly discussed.

\end{minipage}
\end{abstract}
\end{titlepage}
\newpage

\hsize 16.5truecm
\vsize 24.0truecm
\def\dm{$\Delta m^2$\hskip 0.1cm }
\def\dmsqua{$\Delta m^2$\hskip 0.1cm}
\def\sn{$\sin^2 2\theta$\hskip 0.1cm }
\def\snf{$\sin^2 2\theta$}
\def\trna{$\nu_e \rightarrow \nu_a$}
\def\trnm{$\nu_e \rightarrow \nu_{\mu}$}
\def\trns{$\nu_e \leftrightarrow \nu_s$}
\def\trnat{$\nu_e \leftrightarrow \nu_a$}
\def\trnmt{$\nu_e \leftrightarrow \nu_{\mu}$}
\def\trne{$\nu_e \rightarrow \nu_e$}
\def\trnst{$\nu_e \leftrightarrow \nu_s$}
\def\nue{$\nu_e$\hskip 0.1cm}
\def\numu{$\nu_{\mu}$\hskip 0.1cm}
\def\nutau{$\nu_{\tau}$\hskip 0.1cm}

\font\eightrm=cmr8
\def\aprle{\buildrel < \over {_{\sim}}}
\def\aprge{\buildrel > \over {_{\sim}}}
\renewcommand{\thefootnote}{\arabic{footnote}}
\setcounter{footnote}{0}
\baselineskip 18pt
\tightenlines
\vglue 1.5cm
\leftline{\bf 1. Introduction}
\vskip 0.3cm
\indent 
   In the present article we derive exact, 
simple and general 
analytic expression for the probability of matter-enhanced two-neutrino
transitions in a medium. The 
matter-enhanced neutrino transitions provide, as is well-known, at least 
three different varieties of neutrino physics solutions of
the solar neutrino problem \cite{LW78,MS85,RSFP,FCNC1,FCNC2}. 
Such transitions, in particular, 
can play important role in the supernovae dynamics
\cite{SNMSW,SNRSFP} and 
can be at the origin of the observed large 
space velocities of pulsars \cite{PULS}. 
Even in the simplest case
of transitions involving only two weak-eigenstate neutrinos, however,
the system of evolution equations describing the transitions
does not admit, in general, exact solutions and one has to 
rely on numerical methods
to calculate the corresponding transition probabilities.
There are few notable exceptions: the evolution equations can
be solved exactly, for instance, in the case of MSW transitions \cite{LW78,MS85} 
in matter with density which changes linearly 
\cite{SP191,WHAX,DAR} (see also \cite{ZEN}) 
or exponentially \cite{SP200,EXP} 
along the neutrino
path. On the basis of the exact solutions expressed in the two cases
respectively in terms of parabolic cylinder and confluent hypergeometric 
functions, simple expressions for the neutrino transition
probabilities, containing only elementary functions, have been 
obtained
\footnote{\baselineskip 18pt The expression for the average MSW 
neutrino transition probability in the case of linearly varying density
was derived first in ref. \cite{LINEAR} using qualitative
arguments and the Landau - Zener formula \cite{ZEN,LAND} 
for the ``jump'' probability.} \cite{SP191,WHAX,DAR,SP200}. 
The case of exponentially varying density
is especially relevant for the analytic description of the 
solar neutrino transitions in the Sun since according to the
contemporary solar models \cite{SSM} the density decreases approximately
exponentially from the center to the surface of the Sun.
Indeed, it was found that the expression for the average probability
obtained in the exponential density approximation provides a very precise
(and actually, the most precise) description of the MSW transitions
of the solar neutrinos in the Sun \cite{KP88}.

  The expression for the average probability derived in the linear density
approximation 
contains as an integral part the result by Landau and 
Zener for the corresponding ``jump'' probability \cite{ZEN,LAND}. 
It is widely used,
for example, in the studies of the MSW transitions and resonance
spin-flavour precession (RSFP) or conversion \cite{RSFP} 
of the supernova neutrinos
(see, e.g., \cite{SNMSW,SNRSFP,PULS}). 
However, its accuracy in describing the supernova neutrino 
transitions/conversions has never been tested. More specifically,
the oscillating terms present in the relevant transition/conversion
probabilities have always been neglected without 
any control on the validity of this approximation because no
explicit expressions for these terms 
could be derived
\footnote{\baselineskip 18pt One exception is the first article in ref. \cite{SNRSFP} 
where the results of the present study have been used to ensure
that the corresponding oscillating terms are strongly suppressed.}.

    The analytical result for the two-neutrino transition/conversion
probability we obtain in the present study, in addition of being
exact and rather simple, is also general and universal in form. 
It is valid for two-neutrino MSW transitions, RSFP, for matter-
enhanced transitions induced by neutrino flavour-changing neutral
current (FCNC) interactions \cite{FCNC1,FCNC2}, etc. In all these cases the 
neutrino transition/conversion probabilities are shown to be given by
expressions which have one and the same universal structure. Such
an expression was derived first in ref. \cite{SP88osc} for the solar 
neutrino MSW transition probability in the exponential density
approximation.

\vskip 0.3cm
\leftline{\bf 2. The Probability of Two-Neutrino Transitions in a Medium: 
General Results}
\vskip 0.3cm
\indent
   Consider neutrino transitions in a medium, which involve two
weak-eigenstate neutrinos $\nu_{\alpha}$ and $\nu_{\beta}$, 
$\nu_{\alpha}\rightarrow \nu_{\beta}$, 
$\nu_{\alpha}\neq \nu_{\beta} = \stackrel{(-)}{\nu}_{e}, 
\stackrel{(-)}{\nu}_{\mu}, \stackrel{(-)}{\nu}_{\tau}, \stackrel{(-)}\nu_{s}$, 
$\nu_{s}$ being a sterile neutrino
\footnote{\baselineskip 18pt In what follows the term 
``transition'' will be used to denote
the process $\nu_{\alpha}\rightarrow \nu_{\beta}$ of the change of the 
type of the weak-eigenstate neutrino for all possible different cases,
including the case of RSFP.}. We shall assume that 
the transitions are described by a system of 
evolution equations which has (or can be reduced to) the form:
\vskip 0.2truecm
\EQ
i\frac{d}{dt}\left(\begin{array}{c}A_{\alpha}(t,t_0)\\A_{\beta}(t,t_0)\end{array}
\right)=\left(\begin{array}{cc}
-\epsilon(t)  & \epsilon'(t)\\
 \epsilon'(t) & \epsilon(t)
\end{array}\right)
\left(\begin{array}{c}A_{\alpha}(t,t_0)\\A_{\beta}(t,t_0)\end{array}\right)
\label{evol}
\EN
\vskip 0.2truecm
\noindent where $A_{\alpha}(t,t_0)$ ($A_{\beta}(t,t_0)$) is the amplitude
of the probability to find neutrino $\nu_{\alpha}$ ($\nu_{\beta}$) at time $t$
of the evolution of the neutrino system if at time $t_0$ the neutrino
$\nu_{\alpha}$ or $\nu_{\beta}$ has been produced, $t \geq t_0$, 
$\epsilon (t)$ and $\epsilon' (t)$ are real functions of some of the
physical quantities characterizing the medium (density, 
magnetic field strength, etc.)
and the neutrino system in vacuum 
(neutrino energy, mass squared difference, etc.),
and we have omitted the indices $\alpha$ and $\beta$ from 
$\epsilon (t)$ and $\epsilon' (t)$. In writing the evolution equations in
the form (1)  
we have supposed that the effects of neutrino absorption,
creation and of possible neutrino instability (or loss of coherence) 
in the evolution of the neutrino system are negligible. 
This corresponds to a large number of physically interesting cases, in particular,
those discussed in
\cite{LW78,MS85,RSFP,FCNC1,FCNC2,SNMSW,SNRSFP,PULS}.
Under the above assumption the evolution matrix of the system is hermitian and
the equations describing the two-neutrino
transitions in a medium can always be brought (by
phase transformations of the two probability amplitudes) to the form (1)
with real $\epsilon (t)$ and $\epsilon' (t)$. 

   There are two specific types of initial conditions for the system (1) 
relevant to our discussion:
$$A^{0}_{\alpha} = 1,~~~ A^{0}_{\beta} = 0,~~~\eqno(A)$$
\noindent and 
$$A^{0}_{\alpha} = 0,~~~ A^{0}_{\beta} = 1,~~~\eqno(B)$$
\noindent where $A^{0}_{\alpha} = A_{\alpha}(t_0,t_0)$ and 
$A^{0}_{\beta} = A_{\beta}(t_0,t_0)$. If the initial conditions (A) hold,
$A_{\alpha}(t,t_0)$ and $A_{\beta}(t,t_0)$ are probability
amplitudes of the $\nu_{\alpha}$ survival and of the
$\nu_{\alpha}\rightarrow \nu_{\beta}$ transition  
while the neutrinos propagate from the point of their production
to the point of neutrino trajectory reached at time $t$: 
$A_{\alpha}(t,t_0) = A(\nu_{\alpha}\rightarrow \nu_{\alpha})$ and
$A_{\beta}(t,t_0) = A(\nu_{\alpha}\rightarrow \nu_{\beta})$
(we have not indicated explicitly that $A(\nu_{\alpha}\rightarrow \nu_{\alpha})$
and $A(\nu_{\alpha}\rightarrow \nu_{\beta})$ depend on $t$ and $t_0$).
Similarly, if the initial conditions (B) are valid we can write
$A_{\alpha}(t,t_0) = A(\nu_{\beta}\rightarrow \nu_{\alpha})$,
$A_{\beta}(t,t_0) = A(\nu_{\beta}\rightarrow \nu_{\beta})$,
in accordance with the 
interpretation of the probability
amplitudes $A_{\alpha}(t,t_0)$ and $A_{\beta}(t,t_0)$.
In the case of most general initial conditions
$$A^{0}_{\alpha} \neq 0,~ A^{0}_{\beta} \neq 0,~
     |A^{0}_{\alpha}|^2 + |A^{0}_{\beta}|^2 = 1, ~~\eqno(2)$$
 
\noindent the solutions of the system (1) are expressed
as linear combinations of the solutions corresponding to the initial
conditions (A) and (B):
$$A_{\alpha}(t,t_0) = A^{0}_{\alpha} A(\nu_{\alpha}\rightarrow \nu_{\alpha}) +
A^{0}_{\beta} A(\nu_{\beta}\rightarrow \nu_{\alpha}),~~~\eqno(3a)$$
$$A_{\beta}(t,t_0) = A^{0}_{\alpha} A(\nu_{\alpha}\rightarrow \nu_{\beta}) +
A^{0}_{\beta} A(\nu_{\beta}\rightarrow \nu_{\beta}).~~~\eqno(3b)$$

\noindent Evidently, for any of the initial conditions (A), (B) or (2)
one has in the case under study
$$|A_{\alpha}(t,t_0)|^2 + |A_{\beta}(t,t_0)|^2 = 1,~~~\eqno(4)$$
\noindent which implies
$$A(\nu_{\alpha}\rightarrow \nu_{\alpha})
A^{*}(\nu_{\beta}\rightarrow \nu_{\alpha}) =
 - A^{*}(\nu_{\beta}\rightarrow \nu_{\beta})
A(\nu_{\beta}\rightarrow \nu_{\alpha})~.~~ \eqno(5)$$

  We give next several concrete examples 
of matter-enhanced neutrino transitions in a medium 
to which our general results will apply.
In the case of MSW $\nu_{e}\rightarrow \nu_{\mu (\tau)}$
transitions of solar or supernova neutrinos we have \cite{LW78,ResCon}
$$\epsilon (t) = {1\over {2}}~[ {~\Delta m^2 \over{2E}}\cos2\theta -
\sqrt{2} G_{F}N_{e}(t)],
~~\epsilon'(t) = {\Delta m^2 \over{4E}}\sin 2\theta, ~~~\eqno(6)$$

\noindent where $\Delta m^2 = m_{2}^2 - m_{1}^{2}$, 
$m_{1,2}$ being the masses of two neutrinos 
$\nu_{1,2}$ with definite mass
in vacuum, $E$ is the neutrino energy, $\theta$ is the neutrino
mixing angle in vacuum, and $N_{e}(t)$ 
is the electron number density
at the point of neutrino trajectory in the Sun 
or supernova, reached at time $t$. 
For the transitions $\nu_{\mu (\tau)}\rightarrow \nu_{e}$
which can take place in supernovae, 
$\epsilon (t)$ and $\epsilon'(t)$ are also given by (6). 
If the MSW transitions are into sterile
neutrino, $\nu_{e}\rightarrow \nu_{s}$, $N_{e}(t)$ in eq. (6) has to be
replaced by \cite{LANG87} $(N_{e}(t) - {1\over{2}}N_{n}(t))$, where $N_{n}(t)$ is
the neutron number density. In supernovae MSW
$\nu_{\mu (\tau)}\rightarrow \nu_{s}$ transitions can also take place.
In the latter case $N_{e}(t)$ in eq. (6) has to be
substituted by $(- {1\over{2}}N_{n}(t))$.
Finally, for the corresponding transitions involving antineutrinos 
the term with the Fermi constant in $\epsilon (t)$ changes sign.

 Another relevant example is the simplest version of 
neutrino resonance spin-flavour precession, e.g., 
$\nu_{e}\rightarrow \bar{\nu}_{\mu (\tau)}$, 
in the Sun or in supernovae. It is described by (1) with \cite{RSFP}

$$\epsilon (t) = {1\over {2}}~[~{\Delta m^2 \over{2E}} -
\sqrt{2} G_{F}(N_{e}(t) - N_{n}(t))],
~~\epsilon'(t) = \mu_{\nu} B_{\bot}(t). ~~~\eqno(7)$$

\noindent In eq. (7)
$\Delta m^2 = m_{2}^2 - m_{1}^{2} \equiv 
m^{2}(\bar{\nu}_{\mu (\tau)}) - m^{2}(\nu_e)$, $\mu_{\nu}$ is a 
$\nu_{e} - \bar{\nu}_{\mu (\tau)}$ transition magnetic moment
of a dipole type and $B_{\bot}(t)$ is the value of
the component of the solar or supernova magnetic field
perpendicular to the neutrino momentum. The RSFP of the type
$\bar{\nu}_{e}\rightarrow \nu_{\mu (\tau)}$, which was shown in ref.
\cite{SNRSFP}
to be one of the possible mechanisms of supernova shock revival,
is also described by (1) with $\epsilon (t)$ and
$\epsilon'(t)$ given by eq. (7) in which the term containing the
Fermi constant has an opposite sign. For neutrino RSFP in a 
twisting magnetic field
\cite{TWISTB1,TWISTB2} as well as MSW transitions 
or RSFP of supernova neutrinos, which can be responsible 
for the observed large space velocities of pulsars \cite{PULS},
the expressions for $\epsilon'(t)$ 
coincide with those given in eqs. (6) (MSW transitions) and (7) (RSFP),
while the expressions (6) and (7) for $\epsilon(t)$ are modified
(but not essentially from the point of view of the problem 
we are interested in)
by the presence of additional terms
\footnote{In all the indicated cases the neutrinos are assumed 
to be relativistic.}.  

  Finally, the system (1) with
$$\epsilon (t) = {1\over{2}} \sqrt{2} G_{F}N_{e}(t)~[\lambda 
(1 + 2~{N_{n}(t)\over{N_e(t)}}) - 1],
~~\epsilon'(t) = \sqrt{2} G_{F}N_{e}(t)~\lambda' 
(1 + 2~{N_{n}(t)\over{N_e(t)}}), ~~~\eqno(8)$$
  
\noindent where $\lambda$ and $\lambda'$ are real constants, 
$\lambda > 0$, corresponds to matter-enhanced transitions of solar
($\nu_{e}\rightarrow \nu_{\tau}$) or supernova 
($\stackrel{(-)}{\nu}_{e}\rightarrow \stackrel{(-)}{\nu}_{\tau (\mu)}$,
$\stackrel{(-)}{\nu}_{\tau (\mu)}\rightarrow \stackrel{(-)}{\nu}_{e}$)
neutrinos induced by $\nu_e$ flavour-changing and new 
$\nu_e$ and $\nu_{\tau (\mu)}$ flavour-conserving 
but flavour non-symmetric,
neutral current interactions (on the $d-$quark) \cite{FCNC1}. 
Note that in the latter case matter-enhanced
$\stackrel{(-)}{\nu}_{e}\rightarrow \stackrel{(-)}{\nu}_{\tau (\mu)}$
($\stackrel{(-)}{\nu}_{\tau (\mu)}\rightarrow \stackrel{(-)}{\nu}_{e}$)
transitions are possible even if neutrinos have zero mass 
and neutrino mixing is absent in vacuum. 

   Our aim is to derive an exact and general expression, e.g., for the 
$\nu_{\alpha}$ survival probability assuming that the functions
$\epsilon(t)$ and $\epsilon'(t)$ are given
\footnote{\baselineskip 18pt The results we shall obtain can be relevant 
also for the inverse problem, namely, the problem of 
reconstructing $\epsilon(t)$ and/or $\epsilon'(t)$ from data about 
$\nu_{\alpha}$ ($\nu_{\alpha}\rightarrow \nu_{\beta}$) survival (transition) 
probability.}. This is typically the case
in the studies of the effects of the matter-enhanced transitions 
of the solar and supernova neutrinos, indicated above. In these studies
one uses the predictions for the electron and 
neutron number density distributions and for the neutrino energy spectra
provided by the standard solar models and by the 
models of the supernovae. The energy spectrum of solar neutrinos is 
practically solar model independent and is 
known with a relatively high precision \cite{SSM}.
The supernova neutrino energy spectra 
are somewhat model dependent \cite{SNMSW,SNRSFP}.
No direct information exists about the magnetic fields in the solar and
supernovae interiors, and in the studies of the neutrino conversion in which
they are presumed to play important role, one uses 
plausible field configurations (see, e.g., the third article quoted in
ref. \cite{RSFP} and ref. \cite{SNRSFP}). 
The magnetic field strength and $\Delta m^2$
in the case of RSFP, the neutrino mixing angle 
in vacuum $\theta$ and $\Delta m^2$
in the case of MSW transitions, and the constants $\lambda$ and $\lambda'$ in 
the example specified by eq. (8), are often treated as free parameters
to be determined by the physics of the problem being investigated.
We will not specify the form of $\epsilon(t)$ and 
$\epsilon'(t)$ in our further
analysis. Our results will be valid for any 
$\epsilon(t)$ and $\epsilon'(t)$ and, in particular, for 
$\epsilon(t)$ and $\epsilon'(t)$ given in eqs. (6) - (8) and the cases 
discussed above. 
   
 Let us remind the reader that neutrinos go through a resonance point 
reached at time $t_{res}$, $t_0 < t_{res} < t$, 
on their way to the final point of their trajectory
if the function $\epsilon(t)$ changes sign going through 
zero at $t = t_{res}$,
$\epsilon(t_{res}) = 0$, while $\epsilon'(t_{res}) \neq 0$ 
(see the first article quoted in ref. \cite{ResCon} as well as ref. \cite{MS85}
and, e.g., \cite{RSFP,FCNC1,BPet87}). These
conditions are typically fulfilled in most of the cases
of physical interest, in which also one has $\epsilon'(t) \neq 0$ on 
the whole neutrino trajectory, except possibly 
at the trajectory's final point. For 
$\nu_{e}\rightarrow \nu_{\mu (\tau)}$ MSW transitions,
for instance, the resonance condition can be satisfied, as it follows from
eq. (6), if, e.g., $\Delta m^2 > 0$ and $\cos 2\theta >0$, and these two
inequalities will be assumed to hold in our further 
discussion of the MSW effect for solar neutrinos.
 
  It is convenient to introduce the neutrino mixing angle in matter,
$$\sin2\theta_{m}(t) = 
{\epsilon'(t)\over{\sqrt{\epsilon^{2}(t) + \epsilon'^{2}(t)}}},~~~\eqno(9)$$

\noindent and the neutrino matter-eigenstates at time $t$, $\nu^{m}_{1,2}(t)$:
$$|\nu_{\alpha}>~ = |\nu^{m}_{1}(t)>\cos\theta_{m}(t) +
                   |\nu^{m}_{2}(t)>\sin\theta_{m}(t),~~~\eqno(10a)$$
$$|\nu_{\beta}>~ = - |\nu^{m}_{1}(t)>\sin\theta_{m}(t) +
                   |\nu^{m}_{2}(t)>\cos\theta_{m}(t).~~~\eqno(10b)$$
  
\noindent At the resonance point we have: 
$\sin^22\theta_{m}(t_{res}) = 1$.
The states $|\nu^{m}_{1,2}(t)>$, which are also called ``adiabatic'',
are the instantaneous eigenstates of the evolution matrix (Hamiltonian)
in (1) at time $t$, corresponding to the two eigenvalues, 
$E^{m}_{1,2}(t)$, whose
difference is given by
$$E^{m}_{2} - E^{m}_{1} = 2\sqrt{\epsilon^{2}(t) + 
\epsilon'^{2}(t)}.~~~\eqno(11)$$

\noindent Under certain specific conditions (e.g., $\nu_e$ produced in the Sun
at densities $N_{e}(t_0)$ much larger than the resonance density
$N^{res}_{e} = \Delta m^2 \cos 2\theta /(2E\sqrt{2}G_{F})$) the states
$|\nu^{m}_{1,2}(t_0)>$, for example, can practically
coincide with the states $|\nu_{\beta,\alpha}>$. 

 We shall denote by 
$A(\nu^{m}_{i}(t_0) \rightarrow \nu^{m}_{j}(t)) \equiv A^{m}_{ij}(t,t_0)$ 
 the probability amplitudes of the 
$\nu^{m}_{i}(t_0) \rightarrow \nu^{m}_{j}(t)$ transitions, $i,j=1,2$,
which take place when the neutrinos propagate
in the medium. The ``jump''
(or ``level crossing'') probability, $P'$,
$$P' =  |A(\nu^{m}_{1}(t_0) \rightarrow \nu^{m}_{2}(t))|^2
= |A^{m}_{12}|^2 ,~~~ \eqno(12)$$  

\noindent plays a very important role in the neutrino matter-enhanced
transitions in a medium: its value determines the type of the
$\nu_{\alpha (\beta)}\rightarrow \nu_{\beta(\alpha)}$
transition (for $P'\cong 0$ it is adiabatic and nonadiabatic otherwise)
and typically controls the value of the
transition probability 
in a large region of the corresponding parameter space
\footnote{\baselineskip 18pt As we shall see later, one has in the case of interest:
$|A(\nu^{m}_{2}(t_0) \rightarrow \nu^{m}_{1}(t))|^2
= |A(\nu^{m}_{1}(t_0) \rightarrow \nu^{m}_{2}(t))|^2 = P'$.}. 
Moreover, in a wide class of cases one can use
the Landau-Zener result for $P'$ \cite{ZEN,LAND}
or, e.g., its analog derived in the 
exponential density approximation \cite{SP200}. Then $P'$ is determined by the
values of $\epsilon(t)$, $\epsilon'(t)$ and of their derivatives at the
resonance point (see further). Therefore we would like to find an 
exact general expression
for the $\nu_{\alpha (\beta)}$ survival and 
$\nu_{\alpha (\beta)}\rightarrow \nu_{\beta(\alpha)}$ transition
probabilities in terms of the
``jump'' probability. To this end let us express  
the four probability amplitudes 
$A(\nu_{\alpha}\rightarrow \nu_{\alpha (\beta)})$ and 
$A(\nu_{\beta}\rightarrow \nu_{\alpha (\beta)})$
in terms of the amplitudes $A^{m}_{ij}$, $i,j=1,2$. This can be done
using the relations (10a) and (10b):
$$A(\nu_{\alpha}\rightarrow \nu_{\alpha})
=~~A^{m}_{11}c^{0}_{m}c_{m} + A^{m}_{12}c^{0}_{m}s_{m} + 
A^{m}_{21}s^{0}_{m}c_{m} + A^{m}_{22}s^{0}_{m}s_{m}, ~~~\eqno(13a)$$ 
$$A(\nu_{\alpha}\rightarrow \nu_{\beta}) 
= - A^{m}_{11}c^{0}_{m}s_{m} + A^{m}_{12}c^{0}_{m}c_{m} - 
A^{m}_{21}s^{0}_{m}s_{m} + A^{m}_{22}s^{0}_{m}c_{m}, ~~~\eqno(13b)$$ 
$$A(\nu_{\beta}\rightarrow \nu_{\alpha})
= - A^{m}_{11}s^{0}_{m}c_{m} - A^{m}_{12}s^{0}_{m}s_{m} + 
A^{m}_{21}c^{0}_{m}c_{m} + A^{m}_{22}c^{0}_{m}s_{m}, ~~~\eqno(13c)$$ 
$$A(\nu_{\beta}\rightarrow \nu_{\beta})
=~~A^{m}_{11}s^{0}_{m}s_{m} - A^{m}_{12}s^{0}_{m}c_{m} - 
A^{m}_{21}c^{0}_{m}s_{m} + A^{m}_{22}c^{0}_{m}c_{m}, ~~~\eqno(13d)$$ 

\noindent where $c^{(0)}_{m} = \cos\theta_m(t_{(0)})$ and 
$s^{(0)}_{m} = \sin\theta_m(t_{(0)})$.

 For any fixed $t_{0}$ and $t \geq t_{0}$ one has from (4) and (13a) - (13d):
$$|A^{m}_{11}|^2 + |A^{m}_{12}|^2 = 1,~~
|A^{m}_{21}|^2 + |A^{m}_{22}|^2 = 1,~~ 
A^{m}_{11} (A^{m}_{21})^{*} = - (A^{m}_{22})^{*} A^{m}_{12}~.~~~\eqno(14)$$

  We shall prove next that the solutions of the system of equations (1)
corresponding to the initial conditions (A) and (B) satisfy the following
relations:
$$A^{*}(\nu_{\alpha}\rightarrow \nu_{\alpha}) = 
A(\nu_{\beta}\rightarrow \nu_{\beta}),~~
A(\nu_{\beta}\rightarrow \nu_{\alpha}) =
 - A^{*}(\nu_{\alpha}\rightarrow \nu_{\beta})~.~~ \eqno(15)$$

\noindent Indeed
\footnote{\baselineskip 18pt For the special case of 
MSW $\nu_e \rightarrow \nu_{\mu (\tau)}$
transitions, eq. (6), these relations were 
shown to be valid in refs. \cite{SP200,AS3nu}.}, 
using the first equation in (1) to express 
$A_{\beta}(t,t_0)$ as
$$A_{\beta}(t,t_0) = 
{1\over{\epsilon'}}~(\epsilon + i{d\over{dt}})A_{\alpha}(t,t_0),~~~\eqno(16)$$

\noindent and substituting eq. (16) in 
the second equation in (1) we get a second 
order differential equation for $A_{\alpha}(t,t_0)$:
$$\left\{~{d^2\over{dt^2}} - {\dot{\epsilon}'\over{\epsilon'}}~{d\over{dt}}
+ [\epsilon^2 + \epsilon'^2 -i\epsilon~({\dot{\epsilon}\over{\epsilon}} -
{\dot{\epsilon}'\over{\epsilon'}})]\right\}A_{\alpha}(t,t_0) = 0,~~\eqno(17)$$

\noindent where $\dot{\epsilon}^{(')} = {d\over{dt}}\epsilon^{(')}$.
It is easy to show that $A^{*}_{\beta}(t,t_0)$ satisfies exactly 
the same second order differential
equation: this can be checked, e.g., 
by taking the complex conjugate of the two equations in
(1) and by using the second equation written as    
$$ - A^{*}_{\alpha}(t,t_0) = 
{1\over{\epsilon'}}~(\epsilon + 
i{d\over{dt}})A^{*}_{\beta}(t,t_0),~~~\eqno(18)$$
 
\noindent to eliminate $(-A^{*}_{\alpha}(t,t_0))$ from the
first equation. Moreover, the initial conditions (A) lead to 
initial conditions for the solution
$A_{\alpha}(t,t_0)$ of the equation (17), which 
coincide with the initial conditions
for the solution $A^{*}_{\beta}(t,t_0)$, following from conditions (B). 
This fact and the relations (16) and (18) lead to 
\footnote{\baselineskip 18pt Obviously, the proof of eq. (15) presented here relies
on the assumption that $\epsilon'(t) \neq 0$ and that the derivatives
of $\epsilon (t)$ and $\epsilon'(t)$ exist on the neutrino trajectory.
} eq. (15). 
 
  It is easy to convince oneself utilizing eqs. (13a) - (13d) and (15) that
the amplitudes $A^{m}_{ij}$ satisfy analogous relations: 
$$A^{m}_{22} = (A^{m}_{11})^{*},~~~  A^{m}_{21} 
= - (A^{m}_{12})^{*}~.~~~\eqno(19)$$

\noindent These relations can be derived also from 
the systems of evolution equations
for the amplitudes $A^{m}_{11}$, $A^{m}_{12}$ and for the amplitudes
$A^{m}_{21}$, $A^{m}_{22}$ which can be obtained 
from (1) by using (13a) - (13d). It follows from (19), in particular, that:
$$\Phi_{11}(t,t_0) =  -~ \Phi_{22}(t,t_0),~~~
\Phi_{21}(t,t_0) = \pi - \Phi_{12}(t,t_0) ,~~~~\eqno(20)$$

\noindent where $\Phi_{ij}(t,t_0)$ is the phase of the amplitude 
$A^{m}_{ij}(t,t_0)$,
$$arg (A^{m}_{ij}) = \Phi_{ij}~, i,j = 1,2.~~~~\eqno(21)$$

\noindent Thus, eqs. (14) and (19) imply that the four complex amplitudes
$A^{m}_{ij}$, $i,j=1,2$, depend actually only on three real 
functions of the parameters of the problem, which can be chosen to be
the ``jump'' probability $P'= |A^{m}_{12}|^2 = |A^{m}_{21}|^2$
and the phase functions $\Phi_{12}(t,t_0)$ and $\Phi_{22}(t,t_0)$.

   It is not difficult to find now the expression of interest for 
the $\nu_{\alpha}$ survival probability, 
$P(\nu_{\alpha} \rightarrow \nu_{\alpha})$,
assuming the initial conditions (A) are valid:
$P(\nu_{\alpha} \rightarrow \nu_{\alpha}) =
|A(\nu_{\alpha}\rightarrow \nu_{\alpha})|^2$.
Note that due to eqs. (4) and (15) the 
other three relevant probabilities,
$P(\nu_{\beta} \rightarrow \nu_{\beta}) =
|A(\nu_{\beta}\rightarrow \nu_{\beta})|^2$ and 
$P(\nu_{\alpha (\beta)} \rightarrow \nu_{\beta (\alpha)}) =
|A(\nu_{\alpha (\beta)}\rightarrow \nu_{\beta (\alpha)})|^2$,
can all be expressed in terms of $P(\nu_{\alpha} \rightarrow \nu_{\alpha})$.
Using eqs. (12), (14), and (19) - (21)  we get from eqs. (13a):
$$P(\nu_{\alpha} \rightarrow \nu_{\alpha};t,t_0) = 
  \bar{P}(\nu_{\alpha} \rightarrow \nu_{\alpha}) + 
\sum_{r=1}^{4} P^{osc}_{r}(t,t_0),~~~\eqno(22)$$

\noindent where
$$\bar{P}(\nu_{\alpha} \rightarrow \nu_{\alpha}) = \frac {1}{2} + 
(\frac{1}{2} - P')\cos2\theta_m(t_0)\cos 2\theta_m(t),~~~\eqno(23)$$

\noindent is the average probability and
$$P^{osc}_{1}(t,t_0) = \sqrt{P'(1-P')}\cos 2\theta_m(t_0)\sin 2\theta_m(t)
~\cos (\Phi_{12} + \Phi_{22}),~~~\eqno(24a)$$
$$P^{osc}_{2}(t,t_0) = - \sqrt{P'(1-P')}\sin 2\theta_m(t_0)\cos 2\theta_m(t)
~\cos (\Phi_{12} - \Phi_{22}),~~~\eqno(24b)$$
$$P^{osc}_{3}(t,t_0) = - \frac {1}{2} P'\sin 2\theta_m(t_0)\sin 2\theta_m(t)
~(\cos 2\Phi_{12} + \cos 2 \Phi_{22}),~~~\eqno(24c)$$ 
$$P^{osc}_{4}(t,t_0) = \frac {1}{2} \sin 2\theta_m(t_0)\sin 2\theta_m(t)~
  \cos 2\Phi_{22},~~~\eqno(24d)$$
\vskip 0.1cm
\noindent are oscillating terms. 
The functions $\Phi_{12}(t,t_0)$ and 
$\Phi_{22}(t,t_0)$ are responsible for the oscillatory dependence of the
four probabilities  
$P(\nu_{\alpha (\beta)} \rightarrow \nu_{\alpha (\beta)})$ and
$P(\nu_{\alpha (\beta)} \rightarrow \nu_{\beta (\alpha)})$
on the neutrino energy $E$ and/or on the parameters characterizing the
transitions being studied. 

    The result expressed by eqs. (22) - (24d) 
was derived for the probability of MSW transitions (see eq. (6))
of solar neutrinos in the Sun in ref. \cite{SP88osc}, 
utilizing the exact solutions of the system (1)
found in the case of electron number 
density $N_e(t)$ changing
exponentially along the neutrino trajectory 
\footnote{\baselineskip 18pt There are several misprints in 
ref. \cite{SP88osc} (a list was given in ref. \cite{SPJR89}). 
The three most relevant 
are: i) the overall minus sign in the right-hand
side of eq. (40) should be canceled, ii) $\cos 2\theta_m(t_0)$ in eq. 
(41) should read $\sin 2\theta_m(t_0)$, and 
iii) a $\pi$ should be added in the right-hand side of the
relation between the 
$\Phi_{12}(t,t_0)$ and the phase of the
$\nu^{m}_{1}(t_0) \rightarrow \nu^{m}_{2}(t)$ transition 
amplitude, given in footnote 9 ($\Phi_{12}(t,t_0)$ defined in \cite{SP88osc}
is expressed in terms of $arg~(- A^{m}_{12})$ rather than of
$arg~(A^{m}_{12})$). Note that the probability 
amplitudes defined in \cite{SP88osc} 
differ from those considered here by the phase factor
$exp~(i\int_{t_0}^{t}\epsilon (t')dt')$, which is also reflected
in the relation between $\Phi_{12}(t,t_0)$ and 
$arg~(A^{m}_{12})$ derived in \cite{SP88osc}.}.  
Explicit analytic expressions for the ``jump'' probability $P'$
as well as for the phase functions  $\Phi_{12}(t,t_0)$ and 
$\Phi_{22}(t,t_0)$ in the indicated case were also
derived \cite{SP200,SP88osc} 
(see also \cite{AASP92}). This permitted to perform 
a detailed study of the magnitude and the behavior under various averagings 
of the oscillating terms present in the solar neutrino MSW transition
probability (for details see refs. \cite{SP88osc,SPJR89}) 
\footnote{\baselineskip 18pt To our knowledge, the 
case of MSW transitions of solar neutrinos in the Sun  
(exponentially varying density) is
the only nontrivial and physically relevant one 
for which explicit analytic expressions 
for the phase functions  $\Phi_{12}(t,t_0)$ and 
$\Phi_{22}(t,t_0)$ were obtained so far.}. 

  For adiabatic transitions one has $P' \cong 0$ and consequently 
$P^{osc}_{1,2,3}(t,t_0) \cong 0$, while 
$P^{osc}_{4}(t,t_0)$ can be non-negligible.
Thus, $P^{osc}_{1,2,3}(t,t_0)$ can be identified 
as nonadiabatic oscillating terms,
while for $P' = 0$, $P^{osc}_{4}(t,t_0)$ should coincide with the oscillating
term in the case of adiabatic transitions. The expression for the 
latter in terms of the functions $\epsilon(t)$ and $\epsilon'(t)$ 
can be easily derived from (1). It has the form of the term  
$P^{osc}_{4}(t,t_0)$, eq. (24d), which allows to determine
$\Phi_{22}(t,t_0)$ when $P' \cong 0$:
$$ \Phi_{22}^{AD}(t,t_0) = 
\int_{t_0}^{t} \sqrt{\epsilon^{2}(t') + \epsilon'^{2}(t')}dt'.~~~\eqno(25)$$

  In the majority of cases 
explicit expressions for $\Phi_{12}(t,t_0)$ and $\Phi_{22}(t,t_0)$ 
do not exist. Nevertheless, eqs. (22) - (24d) permit to
determine, in particular, some of the conditions 
under which the oscillating terms
give negligible contributions in the probabilities 
$P(\nu_{\alpha (\beta)} \rightarrow \nu_{\alpha (\beta)})$ and
$P(\nu_{\alpha (\beta)} \rightarrow \nu_{\beta (\alpha)})$; they also
allow one to obtain upper limits on these contributions when the latter
are expected to be non-negligible. This can be done 
if the probability $P'$ and the values of $\theta_m(t_0)$ and $\theta_m(t)$,
i.e., the values of the $\epsilon(t)$ and 
$\epsilon'(t)$ in the initial and final 
points of neutrino trajectory, are known. In this case the amplitude of the
oscillations described by each of the terms given in eqs. (24a) - (24d) 
can be calculated.

  In many cases of physical interest one can use the Landau-Zener 
expression for the ``jump'' probability \cite{ZEN,LAND} (see also \cite{LINEAR}):
\vskip -0.1cm
$$P'_{LZ} = e^{-2\pi n_0},~~~\eqno(26)$$
\noindent where
$$4n_0 = 4n(t=t_{res}) = 
2~{{\epsilon'^{2}(t=t_{res})}\over{|\dot{\epsilon}(t=t_{res})|}}~,
\eqno(27)$$  
\vskip 0.2truecm
\noindent is the adiabaticity parameter, which is just the value of the 
adiabaticity function
$$4n(t) = ~{{E^{m}_{2} - E^{m}_{1}}\over{2~|\dot{\theta}_{m}(t)|}}~ =~
 2~{{(\epsilon^{2}(t) + \epsilon'^{2}(t))^{\frac{3}{2}}}\over
{|\epsilon(t) \dot{\epsilon}'(t) - \dot{\epsilon}(t) \epsilon'(t)|}}
~~~\eqno(28)$$
\vskip 0.2truecm
\noindent at the resonance point. In order for a 
given type of transition to be
adiabatic, the inequality $4n(t) \gg 1$ should 
be fulfilled at each point of the
neutrino trajectory. In certain specific cases, 
such as MSW transitions of solar
neutrinos in the Sun, the adiabaticity condition 
$4n(t) \gg 1$ is always satisfied
if it is valid at the resonance point \cite{MS85}, i.e., if the inequality 
$4n_{0} \gg 1$ holds. 

  The Landau-Zener expression for $P'$, eq. (26), was derived assuming that 
$\epsilon(t)$ decreases linearly with time (distance)
while $\epsilon'(t)$
is constant, on the neutrino trajectory. The limits of its applicability
for describing the matter-enhanced neutrino transitions in a medium
are well-known \cite{SP191,SP200,KP88}. 
It reproduces the ``jump'' probability
rather accurately if the resonance region of the transition 
\footnote{This is the region of the neutrino trajectory around 
the resonance point, where $\sin^22\theta_{m}(t) \geq 1/2$.} 
is sufficiently narrow, so that $\epsilon'(t)$ does not change
substantially and the change of $\epsilon(t)$ in it can be well approximated
by a linear function. Expression (26) cannot be used for description
of nonadiabatic transitions when the point of neutrino production
is located in the resonance region. 

  In certain specific cases the analytic expression for the ``jump''
probability, derived for matter (electron, neutron number) 
density changing exponentially along the neutrino trajectory \cite{SP200},
$P'_{exp}$, provides a more accurate description of the matter-
enhanced neutrino transitions in a medium than $P'_{LZ}$. For MSW neutrino
transitions, eq. (6), the exponential density result for $P'$ reads :
$$P'_{exp} = ~{{e^{-2\pi n_{0}(1 - \tan^2\theta)}
          - e^{-2\pi n_{0}(\tan^{-2}\theta - \tan^2\theta)}}
         \over {1 - e^{-2\pi n_{0}(\tan^{-2}\theta - \tan^2\theta)}}} 
                        ~= ~{{e^{-2\pi r_{0}{\Delta m^2\over {2E}}\sin^2\theta}
          - e^{-2\pi r_{0}{\Delta m^2\over {2E}}}}
         \over {1 - e^{-2\pi r_{0}{\Delta m^2\over {2E}}}}}~~.~\eqno(29)$$
\vskip 0.3truecm
\noindent Here $r_{0} = |N(t=t_{res})/\dot{N}(t=t_{res})|$,
$\dot{N}(t) = {d\over{dt}}N(t)$ and $N(t) = N_{e}(t)$, i.e.,
$r_{0}$ is the scale-height of the change of $N_{e}(t)$ in the case of 
interest. The analogous expression for the ``jump'' 
probability corresponding to 
RSFP in a constant magnetic field,
$P'_{exp}(RSFP)$, can formally be obtained from eq. (29) by replacing 
$\Delta m^2$ and $\theta$ by $\Delta m^2/\cos 2\delta$ 
and $\delta$ respectively, 
where $\tan 2\delta = 2\mu_{\nu} B_{\bot}/(\Delta m^2/2E)$, and by 
choosing the appropriate $N(t)$ in the expression for $r_{0}$  
(in the case of eq. (7), for example, $N(t) = N_{e}(t) - N_{n}(t)$).

   Unlike expression (26), the one given by eq. (29) describes correctly, for
instance, the (strongly) nonadiabatic MSW transitions of solar 
neutrinos in the Sun for values of $\sin^22\theta \geq (0.2 - 0.3)$
\cite{SP191,SP200,KP88}. 
In the case of RSFP in a magnetic field which varies along 
the neutrino trajectory, one should use the value of the 
field at the resonance
point, $B_{\bot}(t=t_{res})$, in the expression for $P'_{exp}(RSFP)$, 
i.e., in the definition of $\tan 2\delta$ given above.  
To our knowledge, the accuracy of the descriptions of the RSFP of 
solar and supernova neutrinos,
based on the ``jump'' probabilities $P'_{LZ}$ and $P'_{exp}(RSFP)$,
has never been thoroughly tested for magnetic fields which 
change along the neutrino trajectory.
     
  Let us note that independently of the value of $P'$, the contribution
of the oscillating terms $P^{osc}_{1,2,3,4}(t,t_0)$ in the probabilities 
$P(\nu_{\alpha (\beta)} \rightarrow \nu_{\alpha (\beta)})$ and
$P(\nu_{\alpha (\beta)} \rightarrow \nu_{\beta (\alpha)})$
will be suppressed, as it follows from eqs. (24a) - (24d), if, for instance,
$\sin 2\theta (t_0) \cong 0$ (e.g., $\theta (t_0) \cong \pi /2$) and 
$\sin 2\theta (t) \cong 0$ (e.g., $\theta (t) \cong 0$). 
In the case of adiabatic
transitions we have $P' \cong 0$ and $P^{osc}_{1,2,3}(t,t_0) \cong 0$, while
the oscillating term $P^{osc}_{4}(t,t_0)$ is suppressed provided
$\sin 2\theta (t_0) \cong 0$ or $\sin 2\theta (t) \cong 0$.

  It is quite well-known (see, e.g., \cite{KP88}) that in the limits
of sufficiently large and sufficiently small $\Delta m^2/(2E)$ in the MSW
case, the solar neutrinos oscillate in the Sun as in vacuum. It is instructive to
see how expression (22) for 
$P(\nu_{e} \rightarrow \nu_{e})$ reduces to the vacuum oscillation one 
in these two limits. For the solar neutrino MSW transitions in the Sun one has
$\theta_{m}(t) = \theta$ (see eqs. (6) and (9)) since $N_e(t) = 0$ at 
the surface of the Sun. If
$\Delta m^2/(2E)$ is sufficiently large, we have $4n_{0} \gg 1$  and
$N^{res}_{e} = \Delta m^2 \cos 2\theta /(2E\sqrt{2}G_{F}) \gg N_e(t_0) \geq 
N_e(t')$, $t_0 \leq t' \leq t$, where $N_e(t_0)$ is the electron number density
in the point of $\nu_e$ production in the Sun. Consequently, as it follows
from eqs. (6), (9), (25) and (26) (or (29)), $\theta_m(t_0) \cong 0$,
$P' \cong 0$, 
and $\Phi_{22} \cong \Delta m^2 (t - t_0)/(2E)$. Using the above equalities
and eqs. (22) - (24d) it is easy to convince oneself that
$P(\nu_{e} \rightarrow \nu_{e})$ indeed reduces to the vacuum oscillation 
probability. We get the same result when 
$\Delta m^2/(2E)$ is sufficiently small, so that 
$\cos 2\theta_m(t_0) \cong - 1$ ($P^{osc}_{2,3,4} \cong 0$) and
$4n_{0} \ll 1$, i.e., the solar neutrinos undergo extremely
nonadiabatic transitions. In this case eq. (29) implies
\footnote{\baselineskip 18pt The Landau-Zener expression 
for $P'$, eq. (26), leads to an incorrect result
since it is not valid, in particular, for relatively
small $\Delta m^2/(2E)$ \cite{SP191,SP200,KP88}.}
$P' \cong \cos^2\theta$ and we have \cite{SP88osc} 
$\Phi_{12} + \Phi_{22} \cong \pi + \Delta m^2 (t - t_0)/(2E)$.
Therefore $\bar{P}(\nu_{e} \rightarrow \nu_{e})$ and
$P^{osc}_{1}$ reduce to the average probability and the oscillating term
in the solar neutrino vacuum oscillation probability (see, e.g., \cite{BPet87}):
$\bar{P}(\nu_{e} \rightarrow \nu_{e}) \cong 1 - 1/2 \sin^22\theta$ and
$P^{osc}_{1}\cong 1/2 \sin^22\theta \cos (\Delta m^2 (t - t_0)/(2E))$.  
Note that the nonadiabatic oscillating term $P^{osc}_{1}$ 
converges to the vacuum oscillating term in this case.
 
   We would like to conclude this Section with the 
following remark. In certain cases
of neutrino transitions in a medium one is faced with the problem
of calculating the real part of the product of the two amplitudes
$A^{*}(\nu_{\alpha}\rightarrow \nu_{\alpha})$ and 
$A(\nu_{\alpha}\rightarrow \nu_{\beta})$, arising as an interference term
in the corresponding transition probability
\footnote{\baselineskip 18pt This term appears, for example, 
in the probability of  
combined MSW transitions and long wave length vacuum oscillations
of solar neutrinos in the case of three flavour-neutrino 
mixing \cite{MSWVO}.}.
Using eqs. (13a), (13b) and (19) - (21) it is not difficult to 
find the expression for the indicated product of amplitudes in terms
of $P'$, $\Phi_{12}(t,t_0)$ and $\Phi_{22}(t,t_0)$:
$$Re~\left [ A^{*}(\nu_{\alpha}\rightarrow \nu_{\alpha}) 
A(\nu_{\alpha}\rightarrow \nu_{\beta})\right ] =
-~(\frac{1}{2} - P')\cos2\theta_m(t_0)\sin 2\theta_m(t) 
+~\sqrt{P'(1-P')}\cos 2\theta_m(t_0)$$
$$\times \cos 2\theta_m(t)~\cos (\Phi_{12} + \Phi_{22})
 +~\sqrt{P'(1-P')}\sin 2\theta_m(t_0)\sin 2\theta_m(t)
~\cos (\Phi_{12} - \Phi_{22})~$$ 
$$-~\frac {1}{2} P'\sin 2\theta_m(t_0)\cos 2\theta_m(t)
~(\cos 2\Phi_{12} + \cos 2 \Phi_{22}) 
 ~+~\frac {1}{2} \sin 2\theta_m(t_0)\cos 2\theta_m(t)~
  \cos 2\Phi_{22}.~~\eqno(30)$$

\noindent Obviously, the terms which depend on the phase functions 
$\Phi_{12}(t,t_0)$ and/or $\Phi_{22}(t,t_0)$ are oscillating terms.
Note that the factors multiplying a given oscillating function
($\cos (\Phi_{12} + \Phi_{22})$, etc.) in the expressions for
$P(\nu_{\alpha} \rightarrow \nu_{\alpha};t,t_0)$ and 
$Re~\left [ A^{*}(\nu_{\alpha}\rightarrow \nu_{\alpha}) 
A(\nu_{\alpha}\rightarrow \nu_{\beta})\right ]$, i.e., the oscillation amplitudes,
are different. The result (30) was used in the recent 
study \cite{MSWVO} of the hybrid MSW + vacuum oscillation solution of the solar
neutrino problem. It should be clear that using eqs. (13a) - (13d), (19) and (20) we
can express any product of two of the amplitudes 
$A^{(*)}(\nu_{\alpha}\rightarrow \nu_{\alpha (\beta)})$ and 
$A^{(*)}(\nu_{\beta}\rightarrow \nu_{\beta (\alpha)})$ in terms of 
$P'$, $\Phi_{12}$, $\Phi_{22}$, $\theta_m(t_0)$ and $\theta_m(t)$. 

\vskip 0.4cm
\leftline{\bf 3. MSW Transitions of Solar Neutrinos in the Sun 
  and  the Hydrogen Atom}
\vskip 0.3cm
\indent In the present Section we demonstrate that the second order
differential equation for the amplitude $A(\nu_{e}\rightarrow \nu_{e})$, 
which describes the MSW  
transitions of solar neutrinos in the Sun, coincides in form
in the case of $N_e(t)$ changing exponentially along the neutrino path,
with the Schroedinger equation for the radial part of the non-relativistic
wave function of the hydrogen atom and we comment briefly on this interesting
coincidence.

    Assuming that i) $\nu_{\alpha} \equiv \nu_{e}$, 
$\nu_{\beta} \equiv \nu_{\mu (\tau)}$, ii) the initial conditions (A) are valid
($A_{e}(t,t_0) = A(\nu_{e}\rightarrow \nu_{e})$), and substituting
$\epsilon (t)$ and $\epsilon' (t)$ in eqs. (16) and (17) 
with their expressions given in eq. (6), we obtain from (17) a second order
differential equation for the amplitude
$A(\nu_{e}\rightarrow \nu_{e})$, describing the MSW   
$\nu_{e}\rightarrow \nu_{\mu (\tau)}$ transitions.
According to the contemporary 
solar models \cite{SSM}, $N_e(t)$ changes approximately
exponentially along the trajectory of a solar neutrino moving radially 
towards the surface of the Sun:
$$N_e(t) = N_e(t_0)\exp{\left\{- {{t - t_0}\over{r_0}}\right\}},~~~~~\eqno(31)$$

\noindent where $(t - t_0) \cong d$ is the distance traveled
by the neutrino in the Sun, $N_e(t_0)$ and $r_0$ have been defined earlier,
$r_0 \sim 0.1R_{\odot}$, $R_{\odot} = 6.96\times 10^{5}~km$ being the solar radius.
Introducing the dimensionless variable 
$$Z = ir_0\sqrt{2}G_{F}N_e(t_0)e^{- {{t - t_0}\over{r_0}}},~~~
 Z_0 = Z(t = t_0),~~~\eqno(32)$$

\noindent and making the substitution
$$A_{e}(t,t_0) \equiv A(\nu_{e}\rightarrow \nu_{e}) = 
(Z/Z_0)^{c-a}~e^{ -(Z - Z_0) + i\int_{t_0}^{t}\epsilon (t')dt'}~A'_{e}(t,t_0),
~~~\eqno(33)$$

\noindent we find that the amplitude $A'_{e}(t,t_0)$ satisfies 
\cite{SP200,EXP,SP88osc} the
confluent hypergeometric equation \cite{BE}:
$$\left\{~Z{d^2\over{dZ^2}} + (c - Z)~{d\over{dZ}}
- a \right\}A'_{e}(t,t_0) = 0,~~\eqno(34)$$   

\noindent where \cite{SP88osc} 
$$ a = 1 + ir_0~{{\Delta m^2}\over{2E}}\sin^2\theta,~~~~ 
c = 1 + ir_0~{{\Delta m^2}\over{2E}}.~~~\eqno(35)$$

  The equation (34) coincides in form 
with the Schroedinger
(energy eigenvalue) equation obeyed by the radial part,
$\psi_{kl}(r)$, of the non-relativistic wave function of the
hydrogen atom \cite{CT}, 
$\Psi(\stackrel{\rightarrow}{r}) = {1\over{r}}\psi_{kl}(r)Y_{lm}(\theta',\phi')$,
where $r$, $\theta'$ and $\phi'$ are the spherical coordinates of the electron
in the proton's rest frame,
$l$ and $m$ are the orbital momentum quantum numbers ($m = -l,...,l$),
$k$ is the quantum number labeling (together with $l$) the electron energy
\footnote{The principal quantum number is equal to $(k + l)$ \cite{CT}.}, 
$E_{kl}$ ($E_{kl} < 0$),
and $Y_{lm}(\theta',\phi')$ are the spherical harmonics. To be more precise,
the function $\psi'_{kl}(Z) = Z^{-c/2}~e^{Z/2}~ \psi_{kl}(r)$ satisfies equation
(34), where the variable $Z$ and the parameters $a$ and $c$ 
are in this case related to the physical quantities characterizing the hydrogen atom:
$$Z = 2~{r\over{a_0}}\sqrt{- E_{kl}/E_{I}},~~
a \equiv a_{kl} = l + 1 - \sqrt{- E_{I}/E_{kl}},~~
c \equiv c_{l} = 2(l + 1),~~\eqno(36)$$ 

\noindent where $a_0 = \hbar/(m_ee^2)$ is the Bohr radius and
$E_{I} = m_ee^4/(2\hbar^2) \cong 13.6~eV$ is the 
ionization energy of the hydrogen atom. It is remarkable that 
the behavior of such different physical systems as solar neutrinos
undergoing MSW transitions in the Sun and the non-relativistic 
hydrogen atom are governed by one and the same differential equation.

  The properties of the linearly independent 
solutions of eq. (34), i.e., of the confluent hypergeometric 
functions, $\Phi(a,c;Z)$, as well as their asymptotic series 
expansions, are well-known
\cite{BE}. Any solution of (34) can be expressed as a linear 
combination of two linearly independent solutions of (34),
$\Phi(a,c;Z)$ and $Z^{1 - c}~\Phi(a-c+1,2-c;Z)$, which are 
distinguished from other sets of linearly independent confluent
hypergeometric functions by
their behavior when $Z \rightarrow 0$: 
$\Phi(a',c';Z = 0) = 1$, $a',c' \neq 0, -1, -2,...$, 
$a'$ and $c'$ being arbitrary parameters. Explicit expressions for the
probability amplitudes $A(\nu_{e}\rightarrow \nu_{e})$ and 
$A(\nu_{e}\rightarrow \nu_{\mu (\tau)})$ in terms of the functions
$\Phi(a,c;Z)$ and $\Phi(a-c+1,2-c;Z)$ were derived in \cite{SP88osc,AASP92}.
In the case of MSW transitions of solar neutrinos
($N_e(t) = 0)$ these expressions have an especially simple form: 
they are given by 
the corresponding  vacuum oscillation amplitudes ``distorted'' by the values of the
functions $\Phi(a',c';Z)$ in the initial point of the neutrino trajectory,
$$A(\nu_{e}\rightarrow \nu_{\mu (\tau)})
= {1\over{2}}~\sin2\theta~\left \{\Phi(a-c,2-c;Z_0) - e^{i(t - t_0)
{{\Delta m^2}\over{2E}}}~\Phi(a-1,c;Z_0)~\right \},~~~\eqno(37)$$  

\noindent etc., where $Z_0$, $a$ and $c$ are defined in eqs. (32) and (35).
In the limit $|Z_0| \rightarrow 0$, which corresponds to zero electron number density,
expression (37) reduces to the one for oscillations in vacuum. 

  It is well-known that the requirement of a correct asymptotic behavior
of the wave function $\psi_{kl}(r)$ at large $r$ leads to the quantization
condition for the energy of the electron, $E_{kl}$, in the hydrogen atom \cite{CT}:
$E_{kl} = - E_{I}/(k + l)^2$, $(k + l) = 1,2,...$ ($l = 0,1,2,...,(k+l)-1$).  
Technically, the condition is derived by using the asymptotic series expansion
of the confluent hypergeometric functions in powers of the argument $Z$ \cite{BE}
(one has $Z \rightarrow \infty$ when $r \rightarrow \infty$, see eq. (36)). 
The same asymptotic series
expansion in the case of the solutions describing the 
MSW transitions of solar neutrinos in the Sun (we have $|Z_0| \gtap 520$ 
in this case \cite{SP88osc}) permitted to obtain  
expression (29) for the ``jump'' probability $P'$ \cite{SP200}, 
as well as explicit expressions for
the phase functions $\Phi_{12}(t,t_0)$ and $\Phi_{22}(t,t_0)$ \cite{SP88osc}. 

\vskip 0.4cm
\leftline{\bf 4. Conclusions}
\vskip 0.3cm
\indent We have derived an exact universal analytic expression
for the probability of two-neutrino matter-enhanced
transitions in a medium (MSW, RSFP, induced by neutrino 
FCNC interaction, etc.). The probability is expressed in terms of
three real functions of the parameters characterizing the
neutrino transitions: the ``jump'' probability $P'$ and two phases
(angles) $\Phi_{12}$ and $\Phi_{22}$. The latter are responsible
for the oscillatory dependence of the probability 
on the parameters of the problem being investigated 
(neutrino energy, matter density, 
magnetic field strength, etc.). Although in the majority of cases
of physical interest the two phase functions
$\Phi_{12}$ and $\Phi_{22}$ are not known, the amplitudes of the
oscillations due to $\Phi_{12}$ and $\Phi_{22}$
are functions only of the ``jump'' probability
and the values of the neutrino mixing angle in matter
at the initial and final points of the 
neutrino trajectory, i.e., of quantities which are typically
estimated or calculated in the studies of the neutrino
matter-enhanced transitions. Our results can be used, in particular,
in the investigations of matter-enhanced transitions of solar
and supernova neutrinos. 

   We have noted also that the second order differential equation 
for the MSW probability amplitude of solar $\nu_e$ survival
in the Sun coincides in form, for solar matter
(electron number) density changing exponentially
along the neutrino trajectory, with the Schroedinger
equation for the radial part of the non-relativistic
wave function of the hydrogen atom. The equation is of a
confluent hypergeometric type and the asymptotic 
series expansion of its solutions plays important role 
in the description of both physical systems:
it permits to obtain, for instance, the quantization condition 
for the electron energy levels in the hydrogen atom
and the analog of the Landau-Zener ``jump''probability
for exponentially varying density 
in the case of MSW transitions of solar neutrinos in the Sun.

\vglue 0.4cm
\leftline{\bf Acknowledgements.} 
The author would like to thank Zhenia Akhmedov, the discussions with whom
of the result expressed by eqs. (22) - (24d) and its derivation, 
known to the author since 1989,
prompted the writing of the present letter.
This work was supported in part 
by the EEC grant ERBFMRXCT960090 and by Grant PH-510 from the
Bulgarian Science Foundation. 

\end{document}